\documentclass[aps,pra,amsmath,amssymb,twocolumn,superscriptaddress,showpacs,floatfix,longbibliography]{revtex4-1}

\usepackage{ifpdf}
 \newif\ifpdf
\ifx\pdfoutput\undefined
   \pdffalse
\else
   \pdfoutput=1
   \pdftrue
\fi
\ifpdf
   \usepackage{graphicx}
   \usepackage{epstopdf}
\else
   \usepackage{graphicx}
\fi

\usepackage{bm}
\usepackage{epsfig}
\usepackage[usenames]{color}
\usepackage{soul}
\usepackage{graphicx}
\usepackage{hyperref}

\begin{document}

\title{Approach for describing spatial dynamics of quantum light-matter interaction in dispersive dissipative media}

\author{A.A. Zyablovsky}
\email{zyablovskiy@mail.ru}
 \affiliation{Dukhov Research Institute of Automatics (VNIIA), 22 Sushchevskaya, Moscow 127055, Russia}
\affiliation{Moscow Institute of Physics and Technology, Moscow 141700, Russia}

\author{E.S. Andrianov}
\affiliation{Dukhov Research Institute of Automatics (VNIIA), 22 Sushchevskaya, Moscow 127055, Russia}
 \affiliation{Moscow Institute of Physics and Technology, Moscow 141700, Russia}
 
\author{I.A. Nechepurenko}
\affiliation{Dukhov Research Institute of Automatics (VNIIA), 22 Sushchevskaya, Moscow 127055, Russia}
\affiliation{Moscow Institute of Physics and Technology, Moscow 141700, Russia}
 
\author{A.V. Dorofeenko}
\affiliation{Dukhov Research Institute of Automatics (VNIIA), 22 Sushchevskaya, Moscow 127055, Russia}
\affiliation{Moscow Institute of Physics and Technology, Moscow 141700, Russia}
\affiliation{Institute for Theoretical and Applied Electromagnetics, 13 Izhorskaya, Moscow 125412, Russia}
 
\author{A.~A.~Pukhov}
\affiliation{Dukhov Research Institute of Automatics (VNIIA), 22 Sushchevskaya, Moscow 127055, Russia}
\affiliation{Moscow Institute of Physics and Technology, Moscow 141700, Russia}
\affiliation{Institute for Theoretical and Applied Electromagnetics, 13 Izhorskaya, Moscow 125412, Russia}

\author{A.~P.~Vinogradov}
\affiliation{Dukhov Research Institute of Automatics (VNIIA), 22 Sushchevskaya, Moscow 127055, Russia}
\affiliation{Moscow Institute of Physics and Technology, Moscow 141700, Russia}
\affiliation{Institute for Theoretical and Applied Electromagnetics, 13 Izhorskaya, Moscow 125412, Russia}

\date{\today}

\begin{abstract}
Solving the challenging problem of the amplification and generation of an electromagnetic field in nanostructures enables to implement many properties of the electromagnetic field at the nanoscale in novel practical applications. A first-principles quantum mechanical consideration of such a problem is sufficiently restricted by the exponentially large number of degrees of freedom, and does not allow the electromagnetic field dynamics to be described if it involves a high number of interacting atoms and modes of the electromagnetic field. Conversely, the classical description of electromagnetic fields is incorrect at the nanoscale due to the high level of quantum fluctuations connected to high dissipation and noise levels. In this paper, we develop the framework with a significantly reduced number of degrees of freedom, which describes the quantum spatial dynamics of electromagnetic fields interacting with atoms. As an example, we consider the interaction between atoms placed in a metallic subwavelength groove, and demonstrate that a spontaneously excited electromagnetic pulse propagates with the group velocity. The developed approach may be exploited to describe non-uniform amplification and propagation of electromagnetic fields in arbitrary dispersive dissipative systems.
\end{abstract}


\maketitle

\section{Introduction}
The study of the interaction between light and matter is a key problem in physics ~\cite{1}. Progress in nanotechnologies ~\cite{2,3,4,5} has made it possible to enhance light-matter interaction at the nanoscale. Such an enhancement plays a crucial role for investigating the influence of the electromagnetic environment, such as photonic crystals, metallic and dielectric plasmonic structures, on the atomic dynamics ~\cite{6,7}. In such structures, engineering of the electromagnetic field density of states allows light-matter interaction to be controlled ~\cite{8}. This control enables conditions for excitation and coherent generation of the electromagnetic field to be achieved. This allows devices like distributed feedback (DFB) lasers ~\cite{9,10,11,12,13,14,15,16,17,18,19,20}, nanolasers, and spasers ~\cite{21,22,23,24,25,26,27,28,29,30,31,32,33}, to be created.

Consistent consideration of the dynamics of electromagnetic fields and atoms is based on quantum electrodynamics. The quantum properties of light arise in theory after the procedure of field quantization, which implies that the electromagnetic field is expanded in a series of system eigenmodes ~\cite{6,34,35,36}. With excitation of the mode, the electromagnetic field appears in the entire mode volume. Thus, if it is essential to consider the temporal evolution of the electromagnetic field in a finite volume, then it is necessary to take into account the infinite number of modes with an appropriate phase relation ~\cite{6,34,35,36}. Such a situation arises in the problem of map coherent superposition from one quantum bit to another, by means of an electromagnetic field ~\cite{37,38,39,40}; the study of ultrafast active plasmonics ~\cite{33,41,42,43}; and the study of laser dynamics with pulsed pumping ~\cite{11,16}. In a full quantum mechanical consideration, the increase of the number of modes leads to the exponential increase of the number of degrees of freedom ~\cite{44}. The same takes place when the number of atoms increases. As a result, a first-principles consideration of the problem of the interaction between atoms and modes of electromagnetic field is impossible for many practical applications. It should be underlined that even in the simplest cases, the first-principles quantum mechanical consideration of spatial dynamics is complicated. For example, the problem of the finiteness of the propagation speed of an electromagnetic signal between two atoms (the so-called Fermi problem ~\cite{45}) has a long history and was solved only recently ~\cite{46,47,48,49,50,51,52,53,54}.

Effects related to the quantum nature of the electromagnetic field and atoms in many practical problems can be addressed without the involvement of exact quantum-mechanical calculations. There are mean-field theories describing the dynamics of a finite number of physical values, neglecting quantum-mechanical correlations ~\cite{55,56,57}. Among these theories, the most extensively used are the rate equations and the Maxwell-Bloch equations ~\cite{55,56,57}. The rate equations can be implemented for laser description, accounting for the spontaneous decay of atoms ~\cite{55,56,57}. They are in a good agreement with the experimental data for lasers with high-quality cavities at large timescales when the stationary regime is set. However, the rate equations do not take into account the phase relations between electromagnetic waves; thus, they are not appropriate to describe many important effects in modern physics, e.g. the propagation of an electromagnetic pulse ~\cite{55}. In the Maxwell-Bloch equations, the classical description of an electromagnetic field is used ~\cite{55,56,57}. Whereas this approach inherits the wave equation for the field and the finiteness of propagation speed of the electromagnetic field, it does not take into account the spontaneous decay of atoms ~\cite{55}. To describe the process of the spontaneous decay, the operators of noise are added to the Maxwell-Bloch equations ~\cite{58,59,60,61,62}. For large numbers of photons and atoms that collectively participate in coherent and fluctuation dynamics, operator equations may be translated into c-number equations ~\cite{59}. However, even in this case, the numerical simulation of these equations requires large computational resources ~\cite{33,63,64}. Moreover, this approach is not suitable for nanosize systems, where the number of photons in the cavity and atoms is small ~\cite{59}.

The aim of this work is to develop the method to describe atoms and electromagnetic mode interaction, taking into account the process of spontaneous emission and the finiteness of the propagation speed of electromagnetic fields in dispersive dissipative media. For this purpose, we obtained the equation system in which the number of equations is a quadratic function of the number of modes, and a linear function of the number of atoms. In the case of a full quantum mechanical calculation through master equations for the density matrix, the number of equations increases exponentially with the number of atoms or modes ~\cite{44}. Implementing our approach makes the consideration of quantum systems with a large number of interacting atoms and modes possible. It is shown that the developed formalism correctly describes the propagation of an electromagnetic pulse with the group velocity. We show that the finiteness of the pulse propagation arises from interference between the electromagnetic field of the different modes. We demonstrate that the rate equations describe interference erroneously, and do not correctly describe ultrafast dynamics. Finally, we show that the electromagnetic pulse that is spontaneously emitted by an atom takes the form of a delta function at initial time, because the rate of spontaneous emission in different cavity modes at initial time does not depend on the difference between atom transition frequency and the eigenfrequency of the mode.

\section{Description of interaction between light and matter}
The dynamics of an interacting electromagnetic field and atoms in a Markov approximation is described by the master equation in the Lindblad form ~\cite{44,58}:
\begin{gather}\label{eq1}
\frac{{\partial \hat \rho }}{{\partial t}} =  - \frac{i}{\hbar }\left[ {{{\hat H}_a} + {{\hat H}_\sigma } + \hat V,\hat \rho } \right] +
\\
\nonumber + {\hat L_a}\left[ {\hat \rho } \right] + \hat L_\sigma ^e\left[ {\hat \rho } \right] + \hat L_\sigma ^{ph}\left[ {\hat \rho } \right] + \hat L_\sigma ^{pump}\left[ {\hat \rho } \right], 
\end{gather}
where ${\hat H_a} = \sum\limits_j {\hbar {\omega _j}\hat a_j^ + {{\hat a}_j}} $  is Hamiltonian of the electromagnetic field, after mode decomposition, ${\hat H_\sigma } = \sum\limits_m {\hbar {\omega _m}\hat \sigma _m^ + {{\hat \sigma }_m}} $  is the Hamiltonian of the two-level atoms, and $\hat V = \sum\limits_{j,\,m} {\left( {\hbar {\Omega _{jm}}\hat a_j^ + {{\hat \sigma }_m} + \hbar \Omega _{jm}^*{{\hat a}_j}\hat \sigma _m^ + } \right)}$  is the interaction between modes and atoms in the Jaynes-Cummings form. Here $\hat a_j^ + $  and ${\hat a_j}$  are respectively the creation and annihilation operators of photons in the $j$-th mode, $\hat \sigma _m^ + $  and ${\hat \sigma _m}$  are respectively the raising and lowering operators for transition of the  $m$-th two-level atom, ${\Omega _{jm}}$  is a coupling constant between the photons in the  $j$-th cavity mode and the  $m$-th atom, ${\omega _j}$  is an eigenfrequency of the  $j$-th cavity mode, and finally ${\omega _m}$  is transient frequency of the  $m$-th atom. The term ${\hat L_a}\left[ {\hat \rho } \right] = \sum\limits_j {\frac{{{\gamma _j}}}{2}\left( {2{{\hat a}_j}\hat \rho \hat a_j^ +  - \hat a_j^ + {{\hat a}_j}\hat \rho  - \hat \rho \hat a_j^ + {{\hat a}_j}} \right)} $  ~\cite{44,58} describes dissipation in each $j$-th mode with dissipation rate ${\gamma _j}$, $\hat L_\sigma ^e\left[ {\hat \rho } \right] = \sum\limits_m {\frac{{{\gamma _D}}}{2}\left( {2{{\hat \sigma }_m}\hat \rho \hat \sigma _m^ +  - \hat \sigma _m^ + {{\hat \sigma }_m}\hat \rho  - \hat \rho \hat \sigma _m^ + {{\hat \sigma }_m}} \right)} $  and $\hat L_\sigma ^{ph}\left[ {\hat \rho } \right] = \sum\limits_m {\frac{{{\gamma _\sigma }}}{2}\left( {{{\hat D}_m}\hat \rho {{\hat D}_m} - \hat \rho } \right)} $  corresponding to the energy and phase relaxations with the rates ${\gamma _D}$  and ${\gamma _\sigma }$, respectively ~\cite{44,58}; the term $\hat L_\sigma ^{pump}\left[ {\hat \rho } \right] = \sum\limits_m {\frac{{{\gamma _{pump}}}}{2}\left( {2\hat \sigma _m^ + \hat \rho {{\hat \sigma }_m} - {{\hat \sigma }_m}\hat \sigma _m^ + \hat \rho  - \hat \rho {{\hat \sigma }_m}\hat \sigma _m^ + } \right)} $  describes pumping of a two-level atom at the rate ${\gamma _{pump}}$ ~\cite{44,58}.

Using the identity $\left\langle {\dot A} \right\rangle  = Tr\left( {\dot \rho A} \right)$  and master equation (1), it is possible to derive the closed system of equations on operator average ${D_m} = \left\langle {\hat \sigma _m^ + {{\hat \sigma }_m} - {{\hat \sigma }_m}\hat \sigma _m^ + } \right\rangle $, ${\varphi _{jm}} = \left\langle { - i\hat a_j^ + {{\hat \sigma }_m}} \right\rangle $  and ${n_{jl}} = \left\langle {\hat a_j^ + {{\hat a}_l}} \right\rangle $. To this end, we split correlations between (i) the average values of the number of photons and the population inversion $ {\left\langle {{{\hat n}_{jl}}{{\hat D}_m}} \right\rangle  = \left\langle {{{\hat n}_{jl}}} \right\rangle \left\langle {{{\hat D}_m}} \right\rangle }$, and (ii) the rising and lowering operators of different atoms ${\left\langle {\hat \sigma _m^ + {{\hat \sigma }_{m'}}} \right\rangle  = {\delta _{mm'}}\left( {\left\langle {{{\hat D}_m}} \right\rangle  + 1} \right)/2} $  ~\cite{55}. This results in the following equations:
\begin{gather}
	\frac{{d{n_{jl}}}}{{dt}} =  - \left( {{\gamma _j} + {\gamma _l}} \right){n_{jl}} + i\left( {{\omega _j} - {\omega _l}} \right){n_{jl}} +
	\\
	\nonumber + \sum\limits_m {\left( {{\Omega _{lm}}{\varphi _{jm}} + \Omega _{jm}^*\varphi _{lm}^*} \right)} 
	\\
	\frac{{d{D_m}}}{{dt}} =  - {\gamma _D}\left( {1 + {D_m}} \right) + {\gamma _{pump}}\left( {1 - {D_m}} \right) - 
	\\
	\nonumber - 2\sum\limits_j {\left( {{\Omega _{jm}}{\varphi _{jm}} + \Omega _{jm}^*\varphi _{jm}^*} \right)} 
	\\
	\frac{{d{\varphi _{jm}}}}{{dt}} =  - {\gamma _\sigma }{\varphi _{jm}} + i\left( {{\omega _j} - {\omega _m}} \right){\varphi _{jm}} +
	\\
	\nonumber +  \frac{{\Omega _{jm}^*}}{2}\left( {{D_m} + 1} \right) + \sum\limits_l {\Omega _{lm}^*{n_{jl}}{D_m}} 
\end{gather}
In Eqs. (2) -- (4), ${D_m}$  is the average value of the operator of the population inversion of the  $m$-th atom ~\cite{56,57}, while ${\varphi _{jm}} = \left\langle { - i\hat a_j^ + {{\hat \sigma }_m}} \right\rangle $ is the average value of the operator that describes the interaction between the electromagnetic field in the $j$-th cavity mode and the $m$-th atom. ${n_{jl}}$  is the average value of the operator of the number of photons in the  $j$-th cavity mode when $j=l$ ; and  ${n_{jl}}$ is the average value of the operator that describes the transition of photons from the  $j$-th cavity mode to the  $l$-th cavity mode when $j \ne l$ . This operator arises from the interference between the electromagnetic field in the  $j$-th and  $l$-th cavity modes. Below, we will demonstrate that neglecting these terms (i.e. interference between the electromagnetic field of different cavity modes) results in instant propagation of the electromagnetic field. We will name the variables ${n_{jl}}$  as cross terms.

Note that the rate equations may be obtained from Eqs. (2) -- (4). ~\cite{55}. First, neglecting the phase relations between electromagnetic modes $\left\langle {\hat a_j^ + {{\hat a}_l}} \right\rangle  = {\delta _{jl}}\left\langle {{{\hat n}_{jj}}} \right\rangle $ , it is possible to reduce Eqs. (2) -- (4) to: 
\begin{gather}
	\frac{{d{n_j}}}{{dt}} =  - 2{\gamma _j}{n_j} + \sum\limits_m {\left( {{\Omega _{jm}}{\varphi _{jm}} + \Omega _{jm}^*\varphi _{jm}^*} \right)}  
	\\
	\frac{{d{D_m}}}{{dt}} =  - {\gamma _D}\left( {1 + {D_m}} \right) + {\gamma _{pump}}\left( {1 - {D_m}} \right) - 
	\\
	\nonumber - 2\sum\limits_j {\left( {{\Omega _{jm}}{\varphi _{jm}} + \Omega _{jm}^*\varphi _{jm}^*} \right)}  
	\\
	\frac{{d{\varphi _{jm}}}}{{dt}} =  - {\gamma _\sigma }{\varphi _{jm}} + i\left( {{\omega _j} - {\omega _m}} \right){\varphi _{jm}} +
	\\
	\nonumber +  \frac{{\Omega _{jm}^*}}{2}\left( {{D_m} + 1} \right) + \Omega _{jm}^*{n_j}{D_m}
\end{gather}
where ${n_j} = {n_{jj}}$  is the average value of the number of photons in the  $j$-th cavity mode. These equations have been used to describe the emission properties of micro- and nanolasers ~\cite{65,66,67,68}.

Secondly, in most types of lasers, the rate of transverse relaxation is larger than that of longitudinal relaxation and the decay rate of the number of photons (i.e. ${\gamma _\sigma } \gg {\gamma _D},\,{\gamma _{pump}},{\gamma _j}$ ). In this case, ${\varphi _{jm}}$  can be adiabatically eliminated from Eqs. (5) -- (7) ~\cite{56}:
\begin{gather}
\frac{{d{n_j}}}{{dt}} =  - {\gamma _j}{n_j} +
\\
\nonumber + \sum\limits_m {\frac{{{\gamma _\sigma }{{\left| {{\Omega _{jm}}} \right|}^2}}}{{\gamma _\sigma ^2 + {{\left( {{\omega _m} - {\omega _j}} \right)}^2}}}\left( {2{n_j}{D_m} + {D_m} + 1} \right)} 
	\\
\frac{{d{D_m}}}{{dt}} =  - {\gamma _D}\left( {1 + {D_m}} \right) + {\gamma _{pump}}\left( {1 - {D_m}} \right) - 
\\
\nonumber - 2\sum\limits_j {\frac{{{\gamma _\sigma }{{\left| {{\Omega _{jm}}} \right|}^2}}}{{\gamma _\sigma ^2 + {{\left( {{\omega _m} - {\omega _j}} \right)}^2}}}\left( {2{n_j}{D_m} + {D_m} + 1} \right)} 
\end{gather}
Eqs. (8) and (9) are the rate equations, also known as the balance equations. They can also be derived from energy balance ~\cite{56}. We show below that both Eqs. (5) -- (7) and Eqs. (8) and (9) are incorrect for the description of the propagation of electromagnetic pulses.

\section{Propagation of electromagnetic field between two atoms}
To demonstrate the main features of Eqs. (2) -- (4), and their advantages over the rate equations, we consider two two-level atoms placed in the sub-wavelength groove in metal (Fig. 1a). In the case of groove that is straight along the z-axis with the profile function $y = \zeta \left( x \right)$ , the eigenmodes of such systems may be presented in the form ${\bf{E}}\left( {{\bf{r}},t} \right) = {{\bf{E}}_{k\omega }}\left( {x,y} \right)\exp \left( {ikz - i\omega t} \right)$ , ${\bf{H}}\left( {{\bf{r}},t} \right) = {{\bf{H}}_{k\omega }}\left( {x,y} \right)\exp \left( {ikz - i\omega t} \right)$ , where ${{\bf{E}}_{k\omega }}\left( {x,y} \right)$  and ${{\bf{H}}_{k\omega }}\left( {x,y} \right)$  are determined through the following equation
\begin{gather}
\left( {{\partial ^2}/\partial {x^2} + {\partial ^2}/\partial {y^2} - {\beta ^2}\left( {k,\omega } \right)} \right)\left\{ \begin{array}{l}
{{\bf{E}}_{k\omega }}\left( {x,y} \right)\\
{{\bf{H}}_{k\omega }}\left( {x,y} \right)
\end{array} \right\} = 0
\end{gather}
where $\beta \left( {k,\omega } \right) = \sqrt {{k^2} - {\omega ^2}/{c^2}} $  in a vacuum and $\beta \left( {k,\omega } \right) = \sqrt {{k^2} - \varepsilon \left( \omega  \right){\omega ^2}/{c^2}} $  in the metal (explicit expressions for  ${{\bf{E}}_{k\omega }}\left( {x,y} \right)$ and  ${{\bf{H}}_{k\omega }}\left( {x,y} \right)$ may be found in ~\cite{69}).

\begin{figure}[t]
	\centering
	\includegraphics[width=0.95\columnwidth]{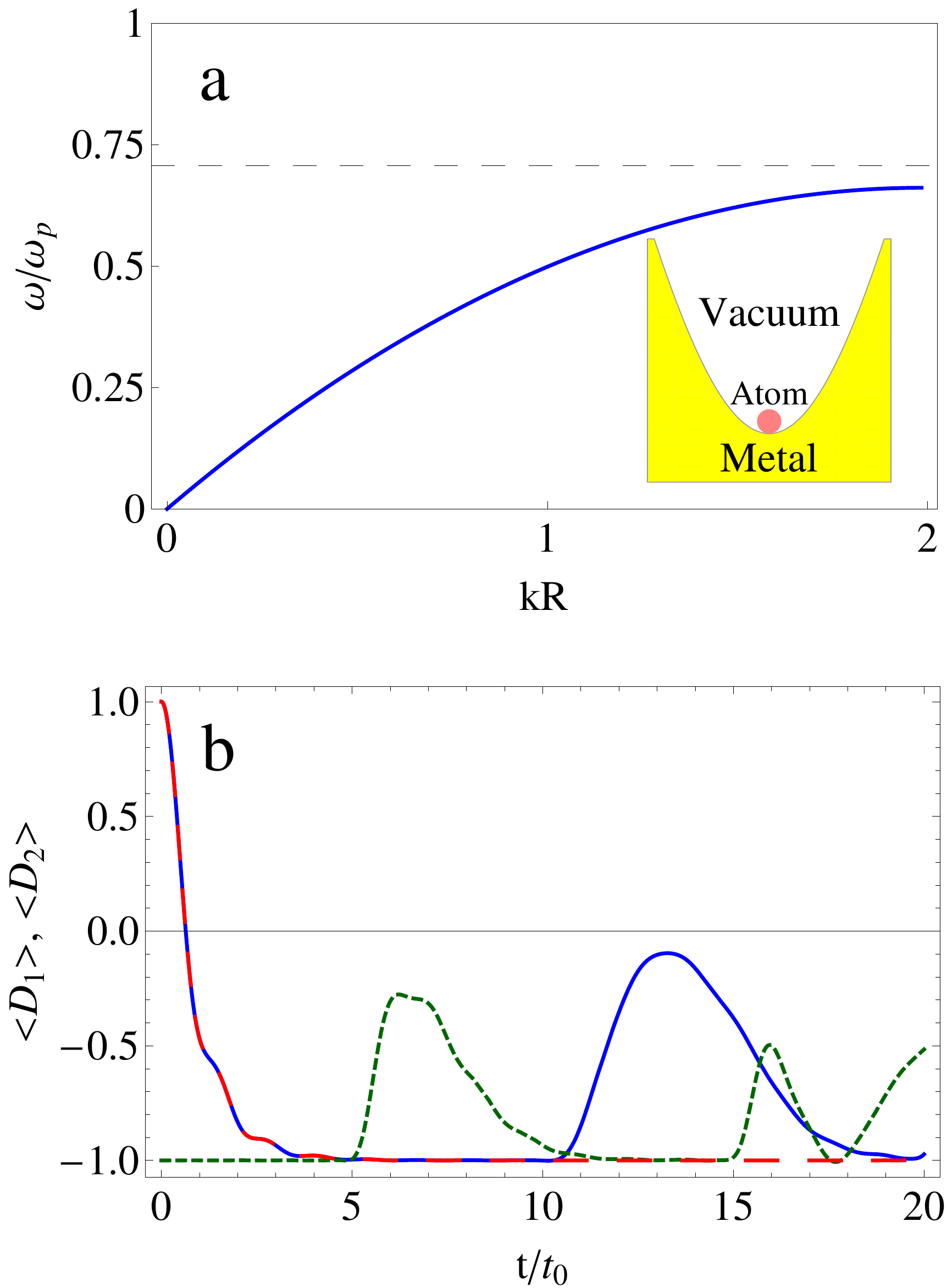}\\
	\caption{(a) Dispersion curve of the eigenmodes of the metallic groove with the profile  $\zeta \left( x \right) =  - A\exp \left( { - {x^2}/{R^2}} \right)$. Inset: schematically illustrated metallic groove. (b) Dependence of population inversion of the first (blue solid line) and second (green dashed line) atoms, and the dependence of the population inversion of the first atom with time in the absence of the second atom (red dot-dashed line) with time obtain from Eqs. (2) -- (4). The distance between the atoms is $l = 500\,{\lambda _\sigma }$. Here ${t_0}$  is equal to ${10^2}{\lambda _\sigma }/{v_g}$ , where ${\lambda _\sigma }$  is a wavelength of the atom transition; ${v_g} = {\left. {\left( {\partial \omega /\partial k} \right)} \right|_{\omega  = {\omega _\sigma }}}$  is group velocity.}\label{fig1}
\end{figure}

After the quantization procedure, the electric and magnetic fields are expressed through creation and annihilation operators for each mode: 
\begin{gather}
{\bf{\hat E}}\left( {{\bf{r}},t} \right) = \sum\limits_j {{A_0}\left( {{k_j}} \right){{\bf{E}}_{{k_j}\omega }}\left( {x,y} \right)} \times
\\	
\nonumber \times \exp \left( {i{k_j}z - i{\omega _j}t} \right){{\hat a}_j} + h.c. 	
\\
{\bf{\hat H}}\left( {{\bf{r}},t} \right) = i\sum\limits_j {{A_0}\left( {{k_j}} \right){{\bf{H}}_{{k_j}\omega }}\left( {x,y} \right)} \times
\\
\nonumber \times \exp \left( {i{k_j}z - i{\omega _j}t} \right){{\hat a}_j} + h.c. 
\end{gather}
The operators ${\hat a_j}$  satisfy the boson commutation relations $\left[ {{{\hat a}_j},\hat a_l^ + } \right] = {\delta _{jl}}\hat 1$ ; the  dimensional constant ${A_0}\left( {{k_j}} \right)$  is determined by the following condition ~\cite{70,71,72,73}:
\begin{gather}
A_0^2\left( {{k_j}} \right)\frac{L}{{8\pi }}\int dxdy ( \partial \left( {\varepsilon \omega } \right)/\partial \omega {{\left| {{{\bf{E}}_{{k_j}\omega }}\left( {x,y} \right)} \right|}^2} + 
\\
\nonumber + {{\left| {{{\bf{H}}_{{k_j}\omega }}\left( {x,y} \right)} \right|}^2} )  = \hbar \omega 
\end{gather}
where $L$  is the groove length.

The interaction between electromagnetic fields and atomic dipole moments in the rotating-wave approximation takes the form of the Jaynes-Cummings Hamiltonian
\begin{gather}
\hat V = \sum\limits_{j,\,m} {\left( {\hbar {\Omega _{jm}}\hat a_j^ + {{\hat \sigma }_m} + \hbar \Omega _{jm}^*{{\hat a}_j}\hat \sigma _m^ + } \right)} 
\end{gather}
with the Rabi frequency
\begin{gather}
{\Omega _{jm}} =  - {{\bf{d}}_{eg}}{A_0}\left( {{k_j}} \right){{\bf{E}}_{{k_j}\omega }}\left( {{x_m},{y_m}} \right)exp\left( {i{k_j}{z_m}} \right)/\hbar  =
\\
\nonumber = {\Omega _{j0}}exp\left( {i{k_j}{z_m}} \right) 
\end{gather}
where ${{\bf{r}}_m} = \left\{ {{x_m},{y_m},{z_m}} \right\}$  is a coordinate of the $m$-th atom.

The decay rate of each mode with wavevector ${k_j}$ is determined by the part of electromagnetic energy inside the metal, and may be evaluated as ~\cite{74,75}:
\begin{gather}
{\gamma _j} = \frac{{{\omega _j}\int\limits_{metal} {dxdy\left( {\varepsilon ''{{\left| {{{\bf{E}}_{{k_j}\omega }}\left( {x,y} \right)} \right|}^2}} \right)} }}{{\int {dxdy\left( {\frac{{\partial \left( {\varepsilon \omega } \right)}}{{\partial \omega }}{{\left| {{{\bf{E}}_{{k_j}\omega }}\left( {x,y} \right)} \right|}^2} + {{\left| {{{\bf{H}}_{{k_j}\omega }}\left( {x,y} \right)} \right|}^2}} \right)} }}
\end{gather}
Let us consider a system of two identical atoms that are placed at a distance of $l$  from each another in the groove with the profile $\zeta \left( x \right) =  - A\exp \left( { - {x^2}/{R^2}} \right)$ , whose dispersion curve is calculated in ~\cite{69}; see Fig. 1a. The first atom at the initial time is in the excited state (i.e. ${D_1}\left( 0 \right) = 1$ ); the second atom at the initial time is in the ground state (i.e. ${D_2}\left( 0 \right) =  - 1$ ); and the number of photons in the cavity is equal to zero. As the operators ${\hat n_{jl}}$  ($j \ne l$ ) and ${\hat \varphi _{jm}}$  change the number of photons in the cavity modes, the average values of these operators ${n_{jl}}$  and ${\varphi _{jm}}$  are equal to zero when the system is in the Fock state (e.g. when the number of photons is equal to zero).

Using Eqs. (2) -- (4), we calculated the dependence of the population inversion of the first and second atoms with time (Fig. 1b).

As follows from Eqs. (2) -- (4), the population inversion of the first atom starts to decay at initial time (Fig. 1b). The population inversion of the second atom is constant and equal to $-1$ until the time is less than propagation time ${\tau _{pr}}$ . After the electromagnetic pulse has reached the second atom, its population inversion begins to increase. The presence of the second atom has no effect on the population inversion of the first atom, until the time is less than two propagation times $2{\tau _{pr}}$ ; see the blue solid and red dot-dashed lines in Fig. 1b. We investigate the dependence of the propagation time ${\tau _{pr}}$  on the distance between atoms, and show that ${\tau _{pr}}$  is a linear function of the distance between atoms (Fig. 2a). This means that the electromagnetic pulse propagates with constant velocity. To investigate the physical nature of this velocity, we change the transition frequency ${\omega _\sigma }$  of the atoms. This leads to a change of the phase and group velocities of the spontaneously emitted electromagnetic pulse at the frequency ${\omega _\sigma }$  (Fig. 2b: blue solid and dashed lines). The numerical simulation of Eqs. (2) -- (4) shows that the population inversion of the second atom begins to increase after the time ${\tau _{pr}} = l/{v_g}$ , where ${v_g} = {\left. {\left( {\partial \omega /\partial k} \right)} \right|_{\omega  = {\omega _\sigma }}}$  is group velocity of the electromagnetic (EM) pulse for the system under consideration (see coincidence of ${\tau _{pr}} = l/{v_g}$, blue line, and the time of atom excitation, bright region, in Fig. 2b). Note that there is an apparent difference between ${\tau _{pr}} = l/{v_g}$  (blue line) and the time of atom excitation (bright region) at the bottom of Fig. 2b. The reason is the rate of increase of the population inversion is determined by the constant of interaction with the EM pulse, which is proportional to the density of states at the atomic transition frequency ${\omega _\sigma }$. Because the density of states is inversely proportional to the group velocity ~\cite{76,77}, when the latter is small, the atom population inversion reaches its maximum value later. We emphasize that neither the phase velocity nor the speed of light in a vacuum affect the atomic population inversion.

\begin{figure}[t]
	\centering
	\includegraphics[width=0.95\columnwidth]{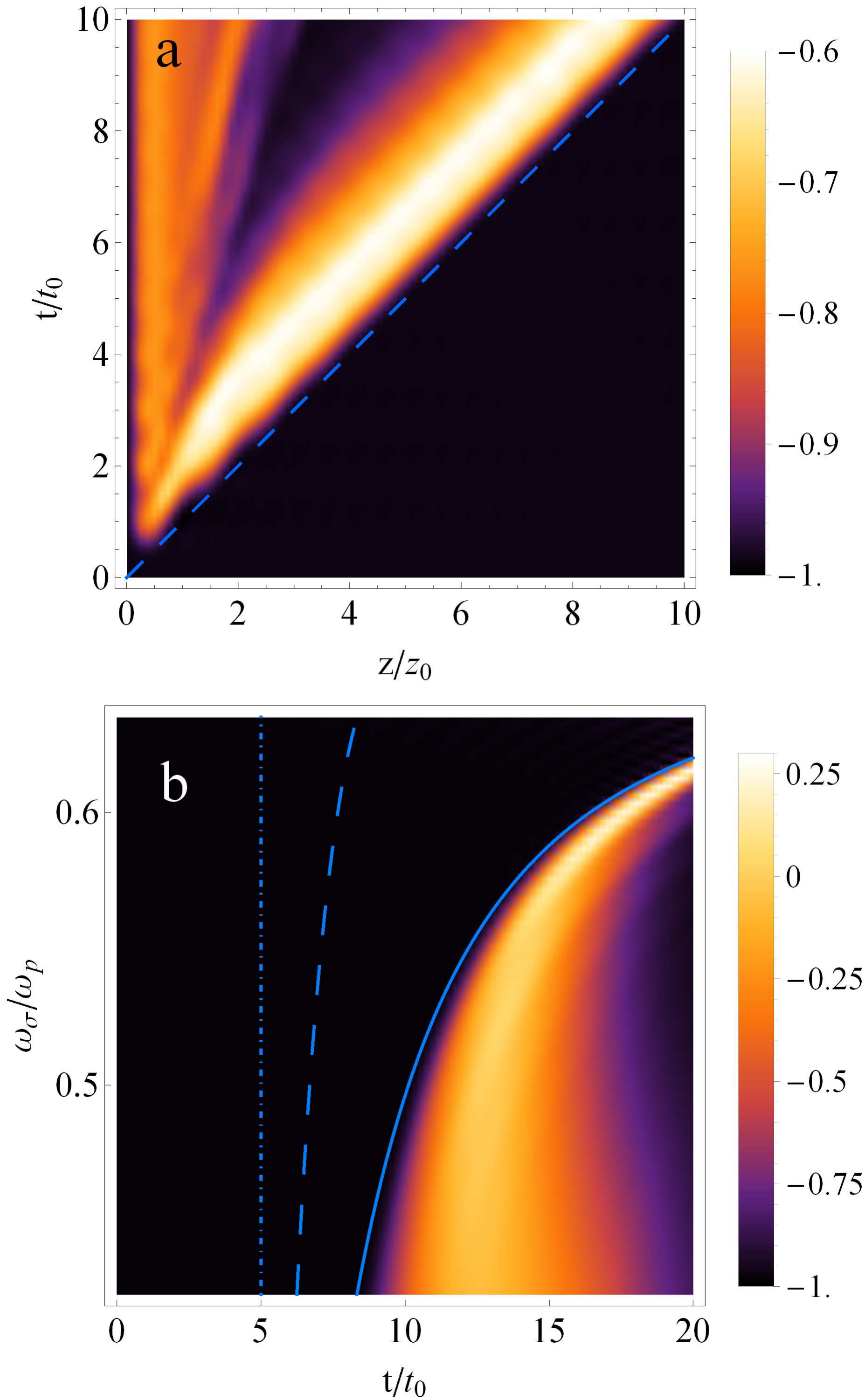}\\
	\caption{(a) Dependence of the population inversion of the second atom on time and distance between atoms obtained from Eqs. (2) -- (4). The blue dashed line is determined by group velocity $z = {v_g}t$ . Here ${z_0}$  is equal to ${10^2}{\lambda _\sigma }$  and ${t_0}$  is equal to ${10^2}{\lambda _\sigma }/{\nu _g}$ . (b) Dependence of the population inversion of the second atom on time and the transition frequency of the atoms obtained from Eqs. (2) -- (4). The solid line is curve  $t = l/{v_g}$ (  ${v_g} = {\left. {\left( {\partial \omega /\partial k} \right)} \right|_{\omega  = {\omega _\sigma }}}$ is group velocity), dashed line is curve $t = l/{v_\phi }$  ( ${v_\phi } = {\left. {\left( {\omega /k} \right)} \right|_{\omega  = {\omega _\sigma }}}$ is phase velocity), and the dashed-dot line is curve $t = l/c$ .}\label{fig2}
\end{figure}

Thus, Eqs.  (2) -- (4) take into account the process of spontaneous emission and the finiteness of the propagation speed of electromagnetic waves. The propagation speed of the electromagnetic signal is equal to the group velocity on the transition frequency of the atom.

\section{Equations without cross terms}
If we assume that $\left\langle {\hat a_j^ + {{\hat a}_l}} \right\rangle  = {\delta _{jl}}\left\langle {{{\hat n}_{jj}}} \right\rangle $ , then Eqs. (2) -- (4) reduce to the equation system (5) -- (7). When Eqs. (5) -- (7) are implemented, the population inversion of the second atom begin to increase with no delay (Fig. 3), and the dynamics of the population inversion of both atoms do not depend on the distance between them.

The reason for the incorrect dynamics description from Eqs. (5) -- (7) can be explained by means of classical electrodynamics. The variables ${\Omega _{jm}}{\varphi _{jm}}$  describe the interaction between the electromagnetic field and the atoms; see Eq. (6). They are proportional to the intensity of the electromagnetic field in the atom location $I$ , which is proportional to the square of the magnitude of the electric field. The electric field can be expanded in the following Fourier series:
\begin{gather}
E\left( {z,t} \right) = \sum\limits_j \left( {{a_j}\exp \left( {i{k_j}z} \right) + a_j^*\exp \left( { - i{k_j}z} \right)} \right) \times
\\
\nonumber \times \exp \left( { - i{\omega _j}t} \right) 
\end{gather}
The amplitudes of the Fourier harmonics ${a_j}$  and $a_j^*$  are classical analogs of the annihilation and creation operators ${\hat a_j}$  and $\hat a_j^ + $ . The intensity of the electromagnetic field is written as
\begin{gather}
I\left( z \right) = \sum\limits_j \sum\limits_l ( {a_j}a_l^*\exp \left( {i\left( {{k_j} - {k_l}} \right)z} \right) + 
\\
\nonumber + a_j^*{a_l}\exp \left( { - i\left( {{k_j} - {k_l}} \right)z} \right) )\exp \left( { - i\left( {{\omega _j} - {\omega _l}} \right)t} \right)  + 
\\
\nonumber + \sum\limits_j \sum\limits_l ( {a_j}{a_l}\exp \left( {i\left( {{k_j} + {k_l}} \right)z} \right) +
\\
\nonumber + a_j^*a_l^*\exp \left( { - i\left( {{k_j} + {k_l}} \right)z} \right) )\exp \left( { - i\left( {{\omega _j} + {\omega _l}} \right)t} \right) 
\end{gather}
The terms in the first sum of Eq. (18) are the classical analogs of  operators of ${\hat n_{jl}} = \hat a_j^ + {\hat a_l}$ . Elements of the sum with $j \ne l$  describe classical mode interference. The terms in the second sum of Eq. (18) oscillate with double frequency. In the rotating wave approximation, these terms are neglected.

Neglecting the cross terms $ {\left\langle {\hat a_j^ + {{\hat a}_l}} \right\rangle  = {\delta _{jl}}\left\langle {{{\hat n}_{jj}}} \right\rangle } $  in Eqs. (2) -- (4) is the equivalent of neglecting of interference terms, because Eq. (18) takes the following form:

\begin{gather}
\tilde I\left( z \right) = 2\sum\limits_j {{{\left| {{a_j}} \right|}^2}}  
\end{gather}

The variable $\tilde I\left( z \right)$  does not depend on the coordinate, which leads to independence of the atomic interaction on the distance between them. This behavior is observed when we use Eqs. (5) -- (7) to describe atomic interaction (Fig. 3).

\begin{figure}[t]
	\centering
	\includegraphics[width=0.95\columnwidth]{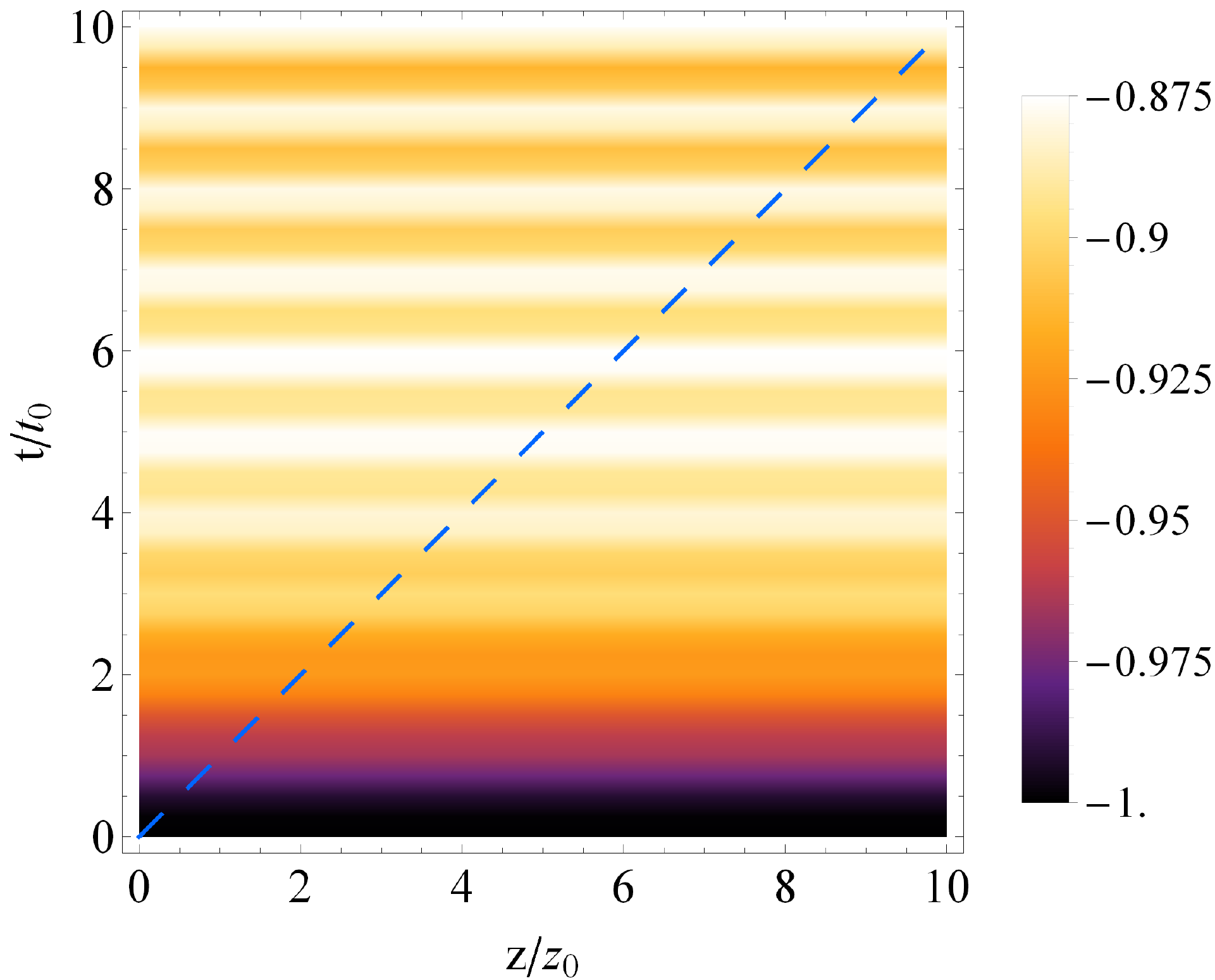}\\
	\caption{Dependence of the population inversion of the second atom on time and distance between atoms obtained from Eqs. (5) -- (7). The blue dashed line is determined by group velocity $z = {v_g}t$ . Here ${z_0}$  is equal to ${10^2}{\lambda _\sigma }$  and   is equal to ${10^2}{\lambda _\sigma }/{v_g}$ , where ${\lambda _\sigma }$ is a wavelength of the atom radiation, ${v_g} = {\left. {\left( {\partial \omega /\partial k} \right)} \right|_{\omega  = {\omega _\sigma }}}$  is group velocity.}\label{fig3}
\end{figure}

Thus, neglecting the cross terms translates Eqs. (2) -- (4) into Eqs. (5) -- (7), and results in instant propagation of the electromagnetic field. As a result, Eqs. (5) -- (7) do not allow for the finiteness of the propagation speed of the electromagnetic field. The rate equations (8) and (9) are derived from Eqs. (5) -- (7) by the adiabatic elimination of the variables of  ${\varphi _{jm}}$. Therefore, the rate equations do not take into account the finiteness of the propagation speed of the electromagnetic field.

\section{Form of the electromagnetic pulse at the initial time of spontaneous decay}
Based on analogy with classical electrodynamics, we determine the following variable
\begin{gather}
I\left( {z,t} \right) = \sum\limits_j \sum\limits_l ( \Omega _l^*\left( z \right){\Omega _j}\left( z \right){n_{jl}}\left( t \right) +
\\
\nonumber + {\Omega _l}\left( z \right)\Omega _j^*\left( z \right)n_{jl}^*\left( t \right) ) 
\end{gather}
which is proportional to the intensity of electromagnetic field at the point $z$ . Here, we used the notation: 
\begin{gather}
{\Omega _j}\left( z \right) = {\Omega _{j0}}\exp \left( {i{k_j}z} \right)
\end{gather}
which is similar to determining the coupling constant between the photons in the cavity modes and the atoms ${\Omega _{jm}}$ ; see Eq. (15).

As follows from Eqs. (2) -- (4), the electromagnetic pulse that was  emitted by the first atom propagates with the group velocity of the EM field and, at the initial time, has a form of a delta function (i.e. the electromagnetic field is different from zero only at the location of the first atom; Figure 4.

\begin{figure}[t]
	\centering
	\includegraphics[width=0.95\columnwidth]{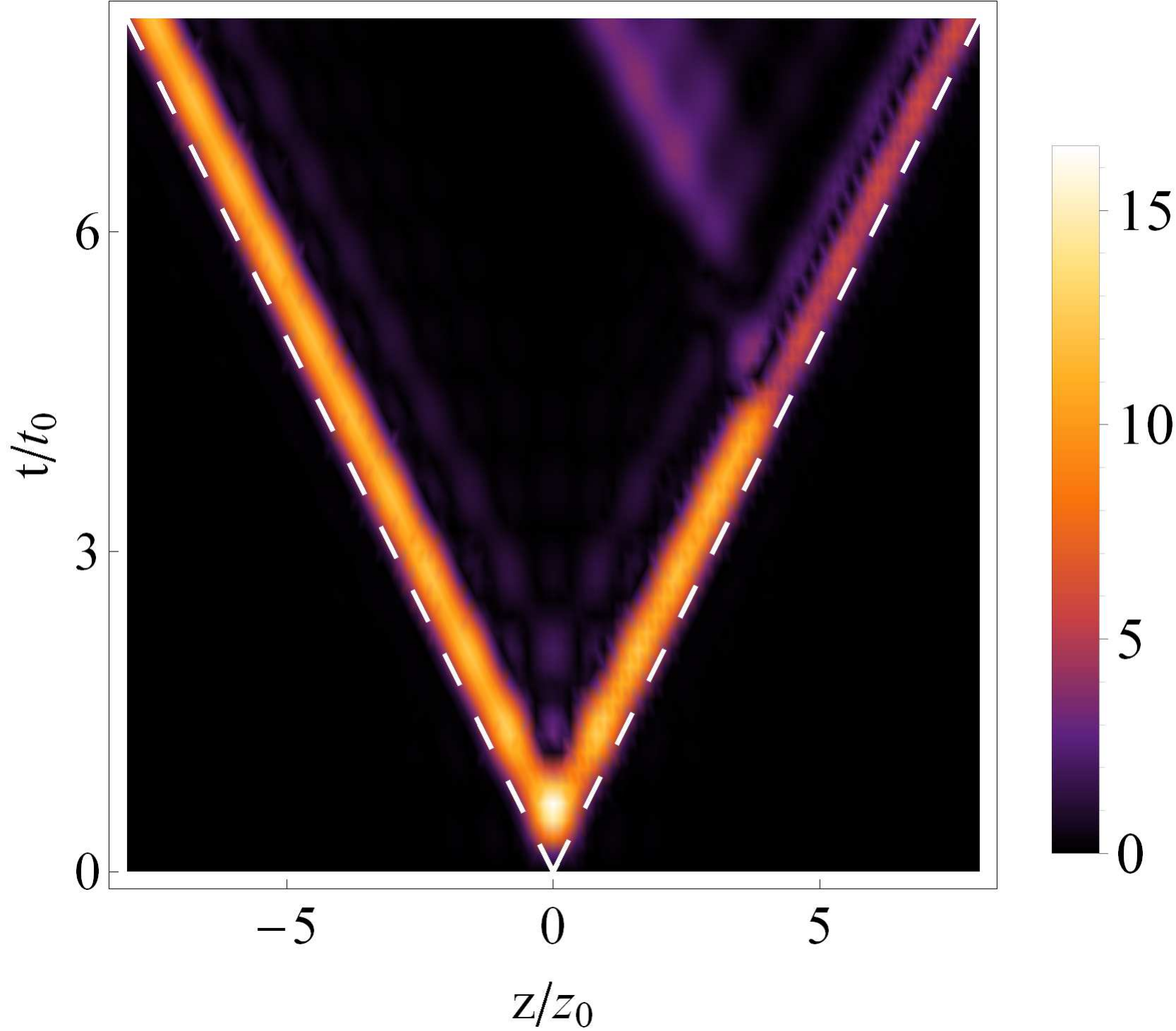}\\
	\caption{Dependence of the intensity of the electromagnetic field in the cavity on time and coordinate obtained from Eqs. (2) -- (4). The white dashed lines are determined by group velocity $z = {v_g}t$ . The first atom is located at $z = 0$ , the second atom is located at $z = 400\,{\lambda _\sigma }$ . Here ${z_0}$  is equal to ${10^2}{\lambda _\sigma }$  and ${t_0}$  is equal to ${10^2}{\lambda _\sigma }/{v_g}$ , where ${\lambda _\sigma }$  is a wavelength of the atom radiation, ${v_g} = {\left. {\left( {\partial \omega /\partial k} \right)} \right|_{\omega  = {\omega _\sigma }}}$  is group velocity.}\label{fig4}
\end{figure}

The Fourier harmonics of the delta function are equal to one another:
\begin{gather}
\delta \left( z \right) = \frac{1}{{2\pi }}\int\limits_{ - \infty }^{ + \infty } {\exp \left( {ikz} \right)dk} 
\end{gather}
Thus, at the initial time of spontaneous decay, the electromagnetic pulse has Fourier harmonics equal to one another. This may be achieved only if all rates of the spontaneous decay ${\Gamma _{sp}}$  in every cavity mode are equal and independent of their eigenfrequencies.

In Eqs. (2) -- (4), the rate of spontaneous decay of the $m$-th atom in the $m$-th cavity mode is proportional to the variable ${\Omega _{jm}}{\varphi _{jm}}$ ; see Eq. (6). Time integration of Eq. (4) results in
\begin{gather}
{\varphi _{jm}}\left( t \right) = \left( {i\left( {{\omega _j} - {\omega _m}} \right) - {\gamma _\sigma }} \right)\int\limits_0^t {{\varphi _{jm}}d\tau }  + 
\\
\nonumber + \frac{{\Omega _{jm}^*}}{2}\int\limits_0^t {\left( {{D_m} + 1} \right)d\tau }  + \sum\limits_l {\Omega _{lm}^*\int\limits_0^t {{n_{jl}}{D_m}d\tau } } 
\end{gather}
At the beginning of the spontaneous decay, the number of photons is equal to zero. As the operators  ${\hat n_{jl}}$ ($j \ne l$) and ${\hat \varphi _{jm}}$  change, the number of photons in the cavity modes, then the average values of these operators ${n_{jl}}$  and ${\varphi _{jm}}$ , are equal to zero when system is in the Fock state (e.g. when the number of photons is equal to zero). As a result:
\begin{gather}
\int\limits_0^t {{\varphi _{jm}}d\tau }  \approx 0
\end{gather}
when the decay time $t$  is less than the characteristic time of the problem.

Therefore, at the initial time, the rate of spontaneous decay does not depend on the difference between the eigenfrequency of the cavity mode and the frequency of the atom transition:
\begin{gather}
{\Omega _{jm}}{\varphi _{jm}}\left( t \right) \approx \frac{{{{\left| {{\Omega _{jm}}} \right|}^2}}}{2}\int\limits_0^t {\left( {{D_m} + 1} \right)d\tau }  = 
\\
\nonumber = \frac{{{{\left| {{\Omega _0}} \right|}^2}}}{2}\int\limits_0^t {\left( {{D_m} + 1} \right)d\tau } 
\end{gather}
where we used the determination of variables ${\Omega _{jm}}$ ; see Eq. (15). As a result, the initial amplitudes of all cavity modes are equal to one another, and the electromagnetic pulse forms a delta function.

Note that in the rate equations (8) and (9), the rate of the spontaneous decay in the cavity mode is proportional to:
\begin{gather}
{\Gamma _{sp}} \sim \frac{{{\gamma _\sigma }{{\left| {{\Omega _{jm}}} \right|}^2}}}{{\gamma _\sigma ^2 + {{\left( {{\omega _\sigma } - {\omega _j}} \right)}^2}}}
\end{gather}
This factor depends on the difference between the eigenfrequency of the cavity mode and the frequency of the atom transition. Therefore, when the rate equations are used, the electromagnetic pulse does not form a delta function at the initial time. This is another reason why the rate equations do not describe the finiteness of the propagation speed of an electromagnetic pulse.

\section{Conclusions}
We have developed a framework to treat the interaction of the electromagnetic field of arbitrary structures and atoms, which takes into account the process of spontaneous decays and the finiteness of the propagation speed of the electromagnetic pulse. As a result, we derive Eqs. (2) -- (4) on operator average ${D_m} = \left\langle {\hat \sigma _m^ + {{\hat \sigma }_m} - {{\hat \sigma }_m}\hat \sigma _m^ + } \right\rangle $ , ${\varphi _{jm}} = \left\langle { - i\hat a_j^ + {{\hat \sigma }_m}} \right\rangle $  and ${n_{jl}} = \left\langle {\hat a_j^ + {{\hat a}_l}} \right\rangle $, which allow us to describe the process of spontaneous decays and the finiteness of the propagation speed of the electromagnetic pulse. Unlike the master equations for the density matrix – in which the number of equations increases exponentially with the number of atoms or modes – in our approach, the number of equations is a quadratic function of the number of the modes and a linear function of the number of the atoms. This opens the possibility to studying open quantum systems consisting of a large number of interacting atoms and modes.

We have demonstrated that when our equations are used, the electromagnetic pulse propagates with the group velocity of the electromagnetic field, and takes the form of a delta function at the initial time of spontaneous decay. It was shown that accounting for the cross terms $\left\langle {{{\hat n}_{jl}}} \right\rangle  = \left\langle {\hat a_j^ + {{\hat a}_l}} \right\rangle $  is necessary for a valid description of the propagation of the electromagnetic pulse in space. Neglecting these terms results in instantaneous propagation of electromagnetic waves in space. 

\section*{Acknowlegments}
We thank Yu. E. Lozovik  for helpful discussions. The work was partially supported by Dynasty Foundation.

\bibliography{ref}

\end{document}